\documentclass[prb,showpacs,twocolumn,amsmath,amssymb,floatfix]{revtex4-1}

\usepackage{array}
\usepackage{graphicx}
\usepackage{bm}
\usepackage[usenames]{color}
\usepackage{dsfont}
\usepackage[utf8]{inputenc}

\newcommand{\beq}{\begin{equation}}
\newcommand{\eeq}{\end{equation}}
\newcommand{\beqa}{\begin{eqnarray}}
\newcommand{\eeqa}{\end{eqnarray}}
\newcommand{\re}{\mathrm{Re}}

\begin{document}

\title{Spontaneous Strains and Gap in Graphene on Boron Nitride}

\author{Pablo San-Jose, \'Angel Guti\'errez, Mauricio Sturla, Francisco Guinea}
\affiliation{Instituto de Ciencia de Materiales de Madrid, Consejo Superior de Investigaciones Cient\'ificas (ICMM-CSIC), Sor Juana In\'es de la Cruz 3, 28049 Madrid, Spain}

\date{\today}

\begin{abstract}
The interaction between a graphene layer and a hexagonal Boron Nitride (hBN) substrate induces lateral displacements and strains in the graphene layer. The displacements lead to the appearance of commensurate regions and the existence of an average gap in the electronic spectrum of graphene. We present a simple, but realistic model, by which the displacements, strains and spectral gap can be derived analytically from the adhesion forces between hBN and graphene. When the lattice axes of graphene and the substrate are aligned, strains reach a value of order 2\%, leading to effective magnetic fields above 100T. The combination of strains and induced scalar potential gives a sizeable contribution to the electronic gap. Commensuration effects are negligible, due to the large stiffness of graphene.
\end{abstract}

\maketitle

\section{Introduction}

Hexagonal Boron Nitride (hBN) has been demonstrated as a promising insulating substrate for graphene. Both systems share the same lattice structure, with a lattice mismatch of $\delta=1.8\%$. hBN  is an insulator with a $\sim 5.2\,\text{eV}$ gap. The electronic carriers in graphene on hBN exhibit very large mobilities\cite{Ponomarenko:NP11,Xue:NM11,Dean:NN10,Yankowitz:14,Kretinin:14}.

The electronic band structure of graphene placed over hBN is being intensively studied, both theoretically \cite{Slawinska:PRB10,Sachs:PRB11,Kindermann:PRB12,Abergel:NJOP13,Mucha-Kruczynski:13,Diez:14,Song:14,Neek-Amal:14,Jung:14,Chen:PRB14,Bokdam:14} and experimentally.\cite{Ponomarenko:N13,Tang:SR13,Yankowitz:NP12,Yu:14,Woods:14} The earliest experiments on different samples showed conflicting results on the existence of an insulating state at the neutrality point. Some experiments\cite{Hunt:S13} suggested the existence of an electronic gap of about $\sim 30\,\text{meV}$, while others do not see any clear evidence of it. \cite{Dean:NN10,Xue:NM11}
There is a growing consensus that inhomogeneous strains in the graphene layer may be the underlying mechanism for gap opening \cite{Jung:14}. While an unstrained and flat graphene monolayer on hBN is expected to be gapless, corrugations and in-plane strains should open a spectral gap.

A recent experiment Re. \onlinecite{Woods:14} strongly suggests the existence of a correlation between the electronic gap and the formation of a peculiar strain pattern on graphene, measured both through conductive atomic force microscopy (AFM) and scanning tunnel microscopy (STM). In the absence of strains (i. e. at large enough rotation angles between the lattice axes, $\theta\gtrsim 1^\circ$), both imaging techniques yield a smoothly varying signal across the sample, following the moir\'e pattern corresponding to the mismatch $\delta$ and the angle $\theta$. As $\theta$ is decreased below $1^\circ$, however, a sudden jump in the AFM and STM patterns occurs. The new AFM pattern is composed of flat hexagonal regions, surrounded by sharp boundaries. It is argued that the hBN crystal creates a rapidly varying adhesion potential landscape \cite{Sachs:PRB11,Neek-Amal:14} to which graphene tries to adapt by deforming. At low angles, within the flat hexagonal regions, graphene is strained to locally compensate for the small rotation and lattice mismatch, thus becoming in registry with the hBN crystal. The accumulated strain is released at the sharp hexagon boundaries. The locally averaged lattice constant, related to the trace of the strain tensor, is measured directly by STM, and is found to differ between hexagonal regions and their boundary by around $2\%$.

 We present here a description of the strains in graphene induced by its adhesion to hBN. We provide an analytical solution for the strains as a function of the twist angle. Using known elastic constants for graphene, and first-principle results for the adhesion potential, we compute the graphene distortion field that globally minimizes the sum of the elastic energy and the adhesion energy. We obtain maximum values for the local expansion of graphene in agreement with the experiment in Ref. \onlinecite{Woods:14}. We also find associated pseudomagnetic fields exceeding 200 T, that are however non-monotonous in the twist angle, and exhibit a global field inversion at a particular angle around $1.5^\circ$. We furthermore characterize the adhesion energy density of the equilibrium graphene solution, and find spatial patterns similar to those in the experiment, with flat hexagonal regions, surrounded by sharp boundaries. Our description of this system provides a simple analytical and quantitative description of most of the features in Ref. \onlinecite{Woods:14}. It may also be used as the basis for an electronic structure computation, and in particular for evaluating the electronic spectral gap associated to these deformations.

The paper is organized as follows. In Sec. \ref{sec:moire}, we set our notation and characterize the geometric moir\'e pattern as a function of lattice mismatch and twist angle. In Sec. \ref{sec:strains} we describe our model for the energetics of adhesion and strain, and write the equilibrium solution for the displacements. We also obtain expressions for the associated pseudomagnetic field. In Sec. \ref{sec:discussion} we plot and discuss the results, including the spectral gap caused by the deformations in Sec. \ref{sec:gap}. Finally, we draw our conclusions in Sec. \ref{sec:conclusion}.

\section{Moir\'e superlattice}
\label{sec:moire}

Graphene and hBN exhibit a $\delta\approx 1.8\%$ lattice mismatch, $a'_0=(1+\delta)a_0$, where $a_0=0.246$ nm and $a'_0=0.251$ nm are the lattice parameters of graphene and hBN, respectively. Thus, a graphene monolayer placed on an hBN crystal will not be in perfect registry, even if their crystallographic axes are perfectly aligned. If both crystals remain strain-free when brought into contact, this results in the formation of a smooth hexagonal moir\'e pattern of period $A_0\approx 14$ nm. If the two crystals are rotated by a relative angle $\theta$, the moir\'e period is reduced. The general form of $A_0$ is
\beq\label{A0}
A_0=|\vec A_1|=|\vec A_2|=\frac{1+\delta}{\sqrt{1+(1+\delta)^2-2(1+\delta) \cos\theta}}a_0
\eeq
where $\vec A_i$ are the superlattice vectors and $|\theta|\leq30^\circ$. This result, and also general expressions for $\vec A_i$, is derived as follows. We write $\vec A_i$, and the corresponding graphene (hBN) lattice vectors $\vec a_i$ ($\vec a'_i$), as the columns of the $2\times 2$ matrices $\bm{A}=(\vec A_1,\vec A_2)=\bm{G}^{-1}/2\pi$, $\bm{a}=(\vec a_1,\vec a_2)=\bm{g}^{-1}/2\pi$ and $\bm{a}'=(\vec a'_1,\vec a'_2)=\bm{g}'^{-1}/2\pi$. By defining the mismatch-plus-rotation transformation $\bm{a}'=\bm{R}\bm{a}$ between the two lattices,
\beq
\bm{R}=(1+\delta)\left(
\begin{array}{cc}
\cos\theta & -\sin\theta \\
\sin\theta & \cos\theta
\end{array}
\right),
\eeq
and by noting that the conjugate momenta  of the moir\'e pattern (rows of matrix $\bm G$) are defined as the mismatch between lattice momenta $\bm{G}=\bm{g}-\bm{g}'$,
\footnote{The moir\'e pattern is a spatial beating pattern, hence the definition of $\bm G$ as the difference of the two spatial frequencies $\bm g$ and $\bm g'$. As in beating waves, this does not necessarily imply periodicity (commensuration) at the atomic level, i.e. at the level of the `carrier waves' of frequency $\bm{g}+\bm{g}'$.}
we find $\bm{A}=\bm{a}\bm{N}=\bm{a}'\bm{N}'$, where $\bm{N}=\bm{a}^{-1}(\mathds{1}-\bm{R}^{-1})^{-1}\bm{a}$ and $\bm{N}'=\bm{a}^{-1}(\bm{R}-\mathds{1})^{-1}\bm{a}$. Eq. (\ref{A0}) follows.

Note that an atomically periodic (commensurate) minimal superlattice is achieved for those values of $\delta$ and $\theta$ that result in fully integer matrices $\bm N=\mathds{1}+\bm N'$. The analysis of the elastic properties that follows, however, are continuum theories that do not rely on precise commensuration, and are generally valid as long as $A_0\gg a_0$.

\begin{figure}
   \centering
   \includegraphics[width=0.62\columnwidth]{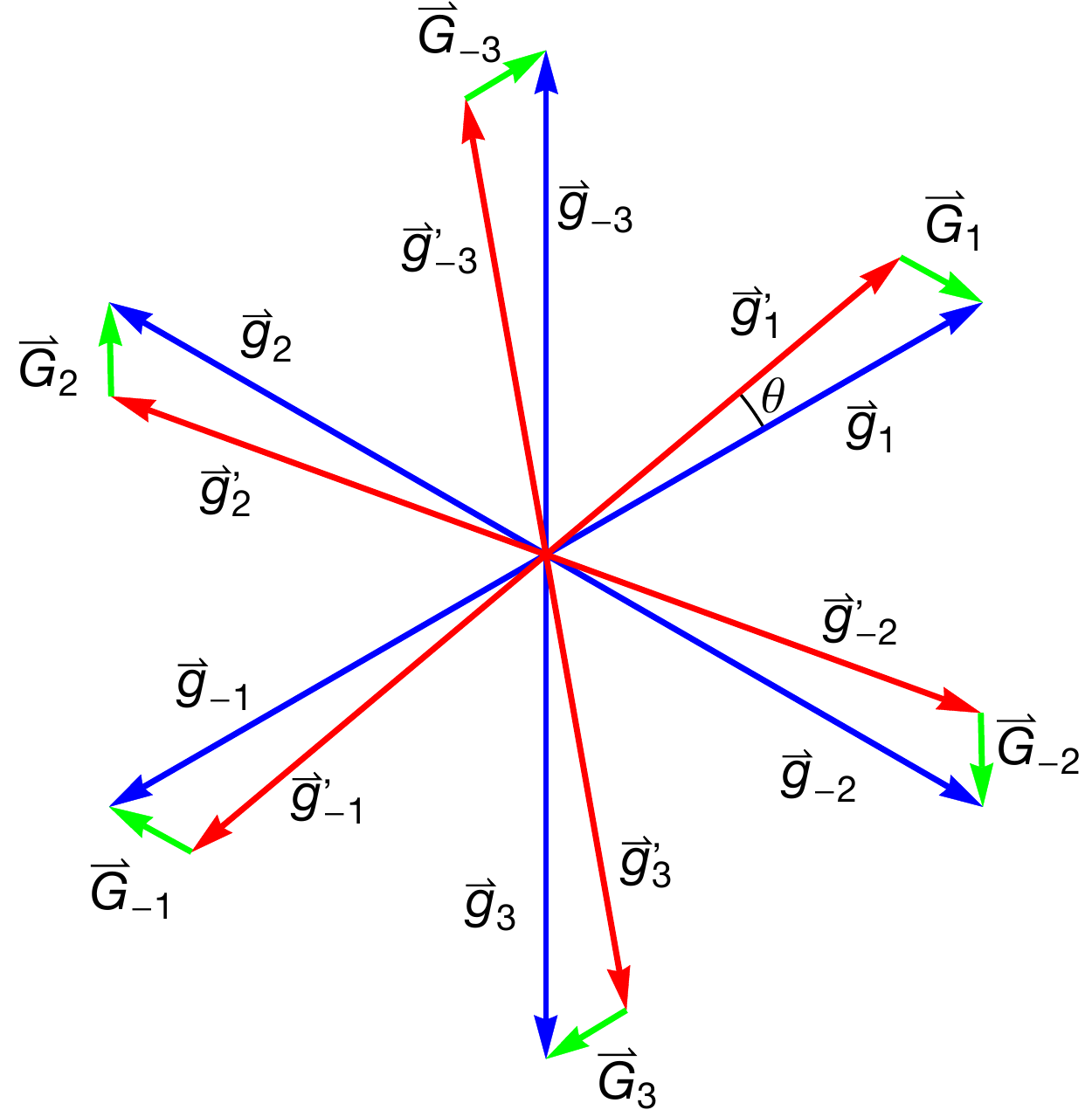}
   \caption{Sketch of the reciprocal lattice vectors $\vec g_j'$ of the hBN lattice (red) and of the graphene lattice $\vec g_j$ (blue). The green vectors $\vec G_j$ describe the moir\'e superlattice, see text. For clarity, the mismatch between the lattice constants of hBN and graphene has been multiplied by 5.}
   \label{fig:sl}
\end{figure}

For later convenience we define here $\vec G_0\equiv 0$, and the momentum ``first star'', which extends the basis $\vec G_{1,2}$ to the six integer combinations thereof that have equal modulus
\beqa \label{firststar}
\vec G_0&=&0\nonumber\\
\vec G_1&=&-\vec G_{-1}=(1,0)\bm G, \nonumber \\
\vec G_2&=&-\vec G_{-2}=(0,1)\bm G, \nonumber \\
\vec G_3&=&-\vec G_{-3}=(-1,-1)\bm G,
\eeqa
We make similar definitions for $\vec g_j$ and $\vec g'_j$, where $j=0,\pm 1,\pm2,\pm3$. A sketch of the reciprocal lattice vectors considered is shown in Fig. [\ref{fig:sl}].

\section{Equilibrium graphene deformation}\label{sec:strains}

The moir\'e superlattice defined in the absence of displacements consists in a smooth spatial variation of the local stacking pattern, which shifts continuously between AA-type (local alignment of both carbons in a unit cell to Boron and Nitrogen), AB-type (Carbon-on-Boron) and BA-type (Carbon-on-Nitrogen). Each of these configurations has a different associated adhesion energy density. Ab-initio calculations \cite{Sachs:PRB11, Caciuc:JPCM12} yield a lower energy for AB stacking, while BA and AA are roughly similar.
The  difference between $\epsilon_{AB}$, $\epsilon_{BA}$ and $\epsilon_{AA}$ adhesion energies in different regions is denoted by
\beqa
\Delta\epsilon_{AB}&=&\epsilon_{AB}-\epsilon_{AA}\nonumber\\
\Delta\epsilon_{BA}&=&\epsilon_{BA}-\epsilon_{AA}\nonumber
\eeqa

These differences in adhesion create in-plane forces in the two crystals. These forces induce distortions which maximize the area of the favorable AB-stacked regions, at the expense of the elastic energy. For a graphene monolayer placed on a thick hBN crystal, it is reasonable to neglect the distortions of hBN. We derive here expressions for the equilibrium graphene displacement field $\vec u(\vec r)$, defined as a minimum of the total energy $U=U_E+U_S$, where $U_E$ is the elastic energy and $U_S$ is the stacking energy (we neglect thermal effects).

\subsection{Elastic energy}

\begin{figure}
   \centering
   \includegraphics[width=0.23\columnwidth]{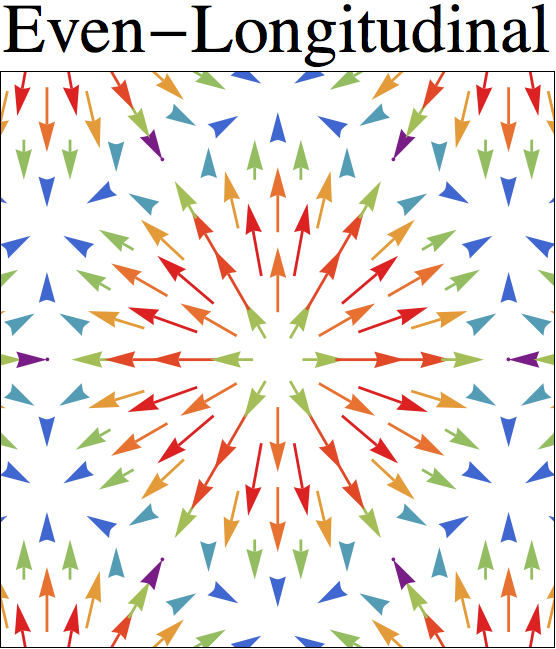}
   \includegraphics[width=0.23\columnwidth]{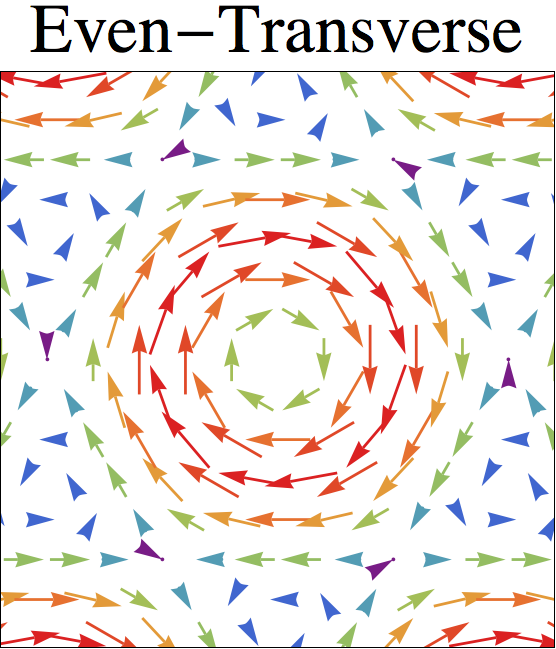}
   \includegraphics[width=0.23\columnwidth]{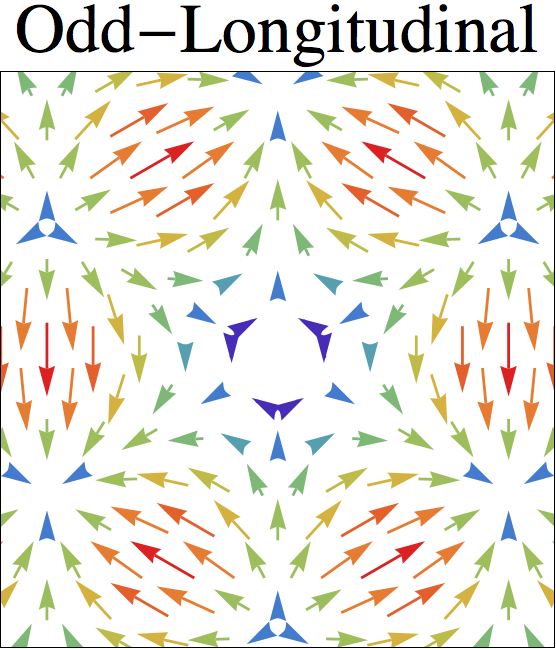}
   \includegraphics[width=0.23\columnwidth]{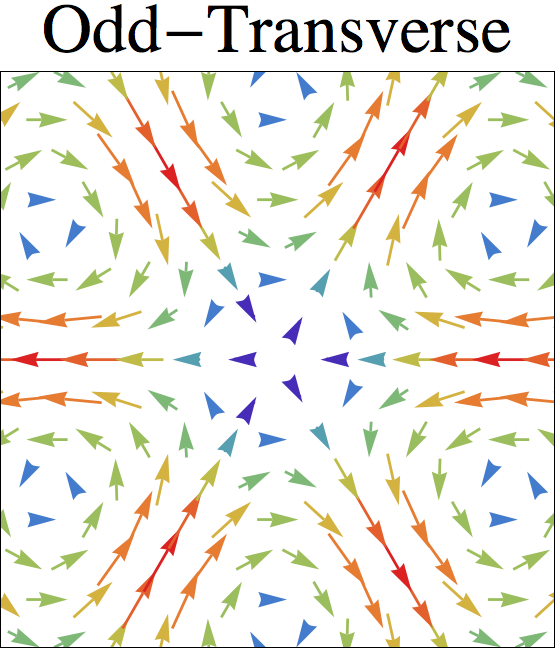}
   \caption{Different types of C${}_3$-symmetric distortions, depending on their longitudinal/transverse character, and their parity respect to the origin.}
   \label{fig:utypes}
\end{figure}

The elastic energy $U_E$ per unit cell of a graphene deformation $\vec u(\vec r)$ that is smooth on the atomic spacing is given by continuum elasticity theory,
\[
U_E=\frac{1}{N}\int_{\bm A} \frac{1}{2}\left[
2\mu \,\mathrm{Tr}(\bm{u}^2)+\lambda \left(\mathrm{Tr} \,\bm{u}\right)^2\right] d^2r ,
\]
where the integral covers a deformation supercell, assumed equal to the moir\'e supercell $\bm A$, which contains $N$ graphene unit cells. Here $\bm u=u_{ij}=\frac{1}{2}(\partial_i u_j+\partial_j u_i)$ is the strain, and $\lambda\approx 3.5~ \mathrm{eV/\AA^2}$ and $\mu\approx 7.8~\mathrm{eV/\AA^2}$ are the Lamé factors for graphene.
Next, we expand the deformation in harmonics $\vec u_{\vec q}=\vec u_{-\vec q}^*$
\beq\label{uharmonics}
\vec u (\vec r)=\sum_{\vec q} \vec u_{\vec q} e^{i\vec q\vec r}
\eeq
Note that, if we assume C${}_3$-symmetric deformations, its harmonics are related by $2\pi/3$-rotations. Taking this into account, we may write all  possible distortions as a combination of four pure classes, see Fig. \ref{fig:utypes}. These are either even or odd respect to a given origin $\vec r_0$, depending on whether $\vec u(\vec r-\vec r_0)=\mp\vec u(-[\vec r-\vec r_0])$ (imaginary or real harmonics if $\vec r_0=0$). They may also be purely longitudinal or purely transverse, depending on whether $\vec u_{\vec q}$ is parallel or perpendicular to $\vec q$.

In terms of the $\vec u_{\vec q}$ harmonics, we may diagonalize $U_E$,
\[
U_E=\frac{1}{2}\sum_{\vec q}\vec u_{-\vec q}\bm{W}_{\vec q}\vec u_{\vec q}
\]
The dynamic matrices $\bm W_{\vec q}$ read
\[
\bm{W}_{\vec q}=(B\det \bm{a})\bm{W}^\parallel_{\vec q}+(\mu\det \bm{a})\bm{W}^\perp_{\vec q},
\]
where $\det \bm a$ is the area of the graphene unit cell, $B=\lambda+2\mu\approx 21.6~\mathrm{eV/\AA^2}$ is graphene's bulk modulus, and
\beq
\bm{W}^{\parallel}_{\vec q}=\left(\begin{array}{cc}
q_x^2 & q_x q_y\\
q_x q_y & q_y^2
\end{array}\right), \hspace{.2 cm}
\bm{W}^{\perp}_{\vec q}=\left(\begin{array}{cc}
q_y^2 & -q_x q_y\\
-q_x q_y & q_x^2
\end{array}\right)
\eeq
They satisfy $\bm W_{\vec q}=\bm W_{-\vec q}=\bm W_{\vec q}^\mathrm{T}$. Note that purely transverse (longitudinal) distortions have only elastic energy contributions from $\mu \bm{W}_{\vec q}^{\perp}$ ($B \bm{W}_{\vec q}^{\parallel}$).

\subsection{Adhesion potential}

\begin{figure}
   \centering
   \includegraphics[width=0.32\columnwidth]{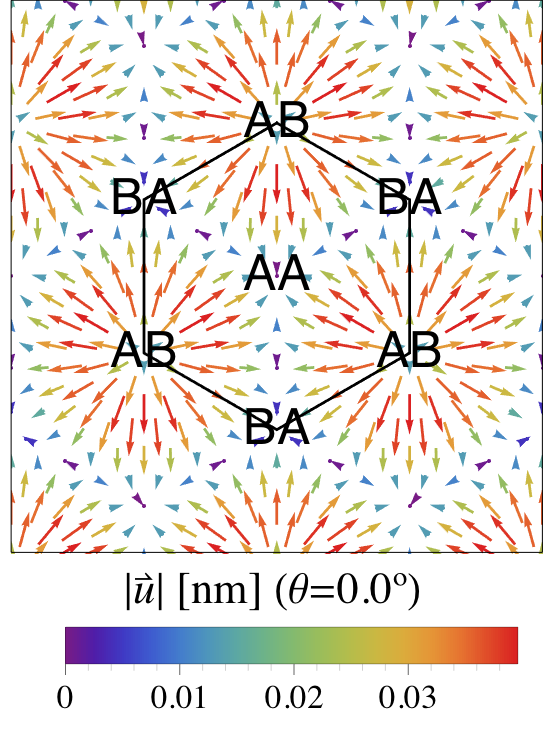}
   \includegraphics[width=0.32\columnwidth]{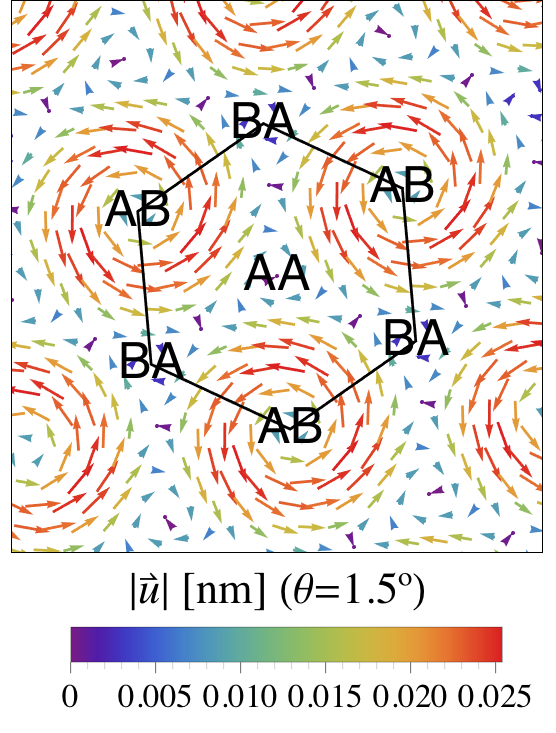}
   \includegraphics[width=0.32\columnwidth]{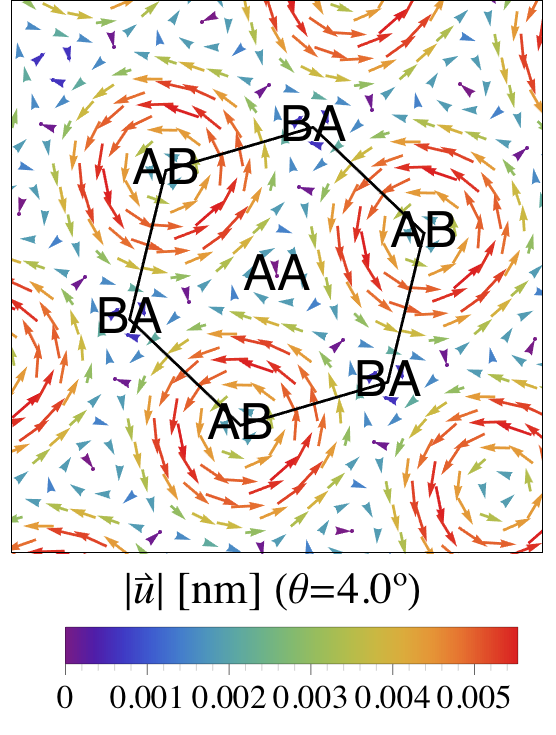}
   \caption{Displacement field $\vec u$ in real space for rotation angle $\theta=0$ (left), $\theta=1.5^\circ$ (center) and $\theta=4^\circ$ (right). Spatial positions are normalized to the moir\'e period $A_0$.}
   \label{fig:u}
\end{figure}

\begin{figure}
   \centering
   \includegraphics[width=0.32\columnwidth]{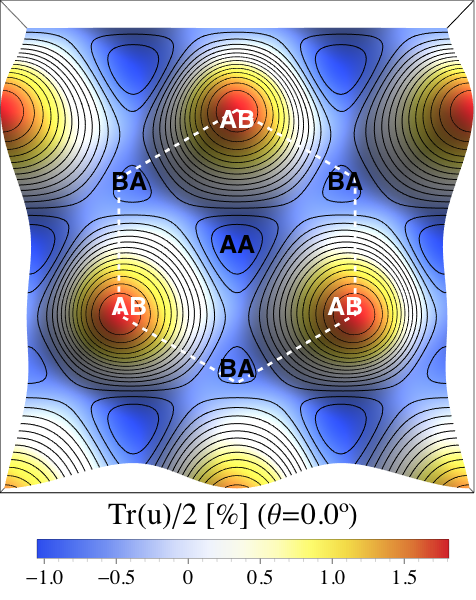}
   \includegraphics[width=0.32\columnwidth]{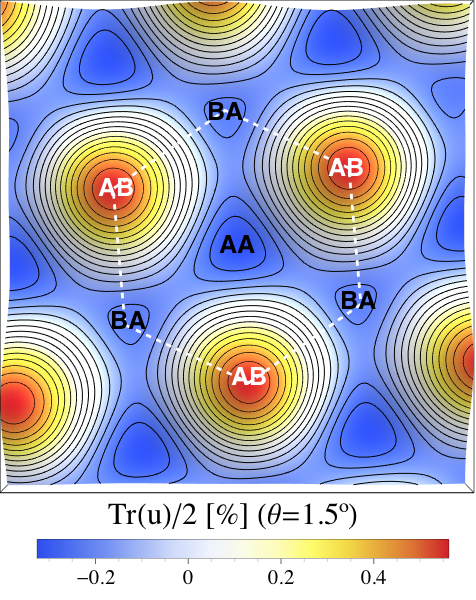}
   \includegraphics[width=0.32\columnwidth]{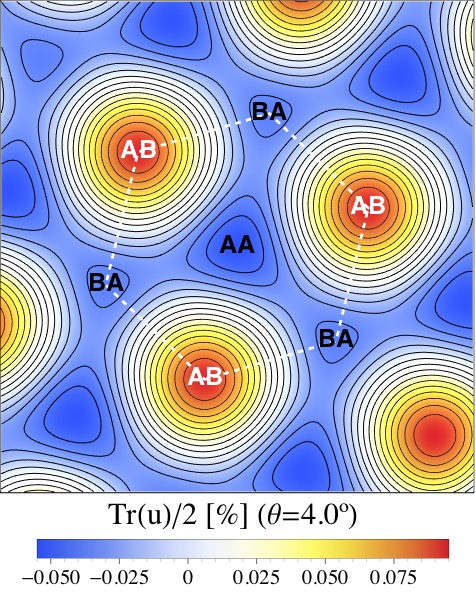}
   \caption{Relative local expansion $\frac{1}{2}\mathrm{Tr}\bm u$ in real space for rotation angle $\theta=0$ (left), $\theta=1.5^\circ$ (center) and $\theta=4^\circ$ (right). Large values of the strain are obtained for $\theta=0$.}
   \label{fig:Tru}
\end{figure}

\begin{figure}
   \centering
   \includegraphics[width=0.32\columnwidth]{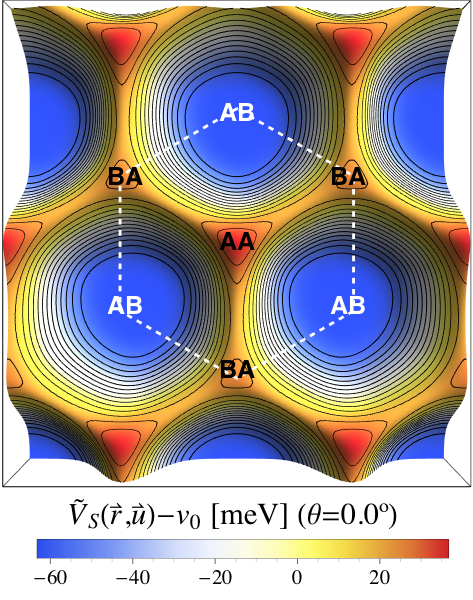}
   \includegraphics[width=0.32\columnwidth]{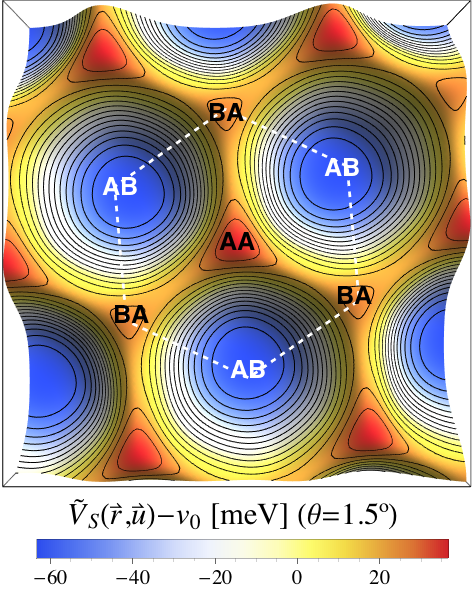}
   \includegraphics[width=0.32\columnwidth]{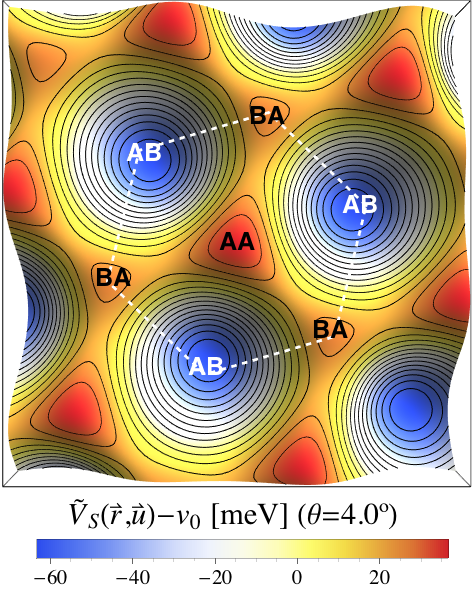}
   \caption{Adhesion energy density $\tilde V_S[\vec r,\vec u(\vec r)]$  in real space, relative to the average adhesion $v_0$, for rotation angle $\theta=0$ (left), $\theta=1.5^\circ$ (center) and $\theta=4^\circ$ (right).}
   \label{fig:Ad}
\end{figure}

\begin{figure}
   \centering
   \includegraphics[width=0.75\columnwidth]{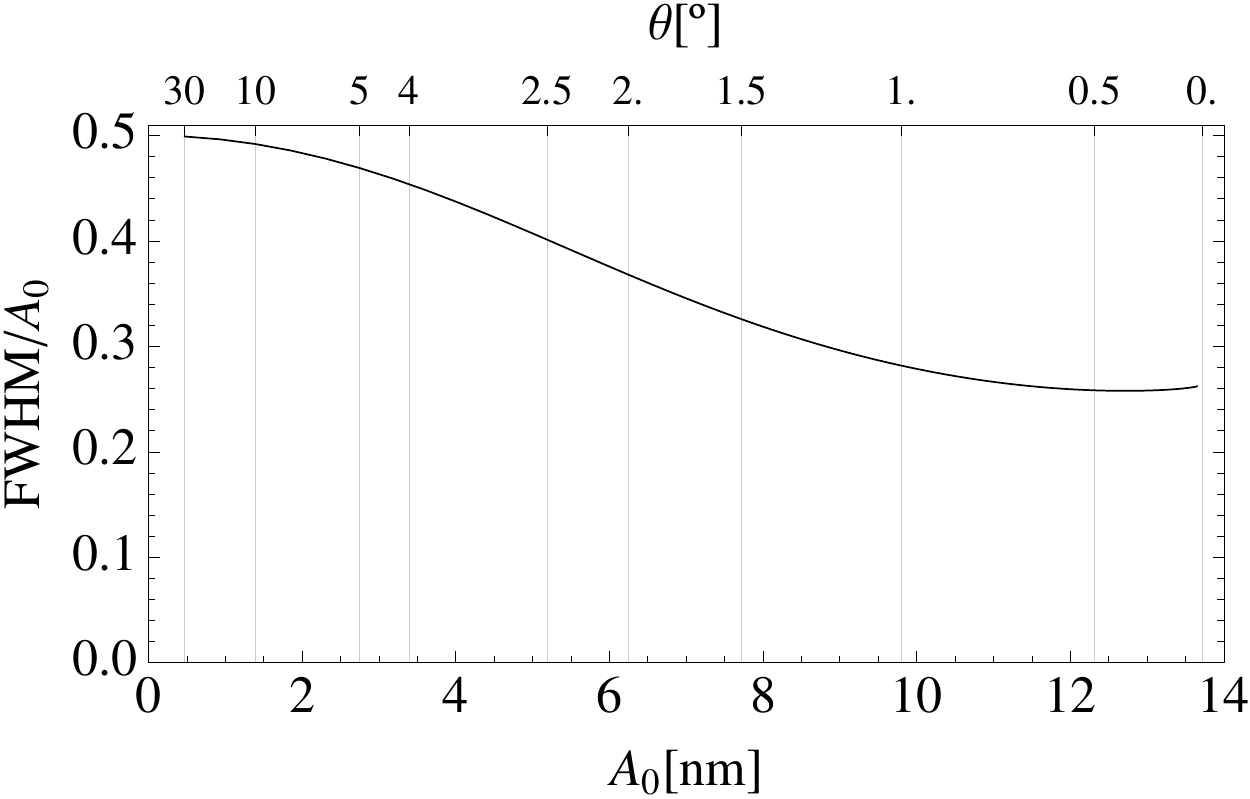}
   \caption{Normalized full width at half maximum (FWHM) of the hexagonal boundaries in the adhesion energy density, see Fig. \ref{fig:Ad}(left), as a function of rotation angle $\theta$, or moir\'e period $A_0$.}
   \label{fig:FWHM}
\end{figure}

We next consider the periodic adhesion potential created by the hBN crystal on the graphene lattice. The simplest model for this potential $V_S(\vec r)$ (``first star'' model), is parameterized by $\Delta\epsilon_{AB}$ and $\Delta\epsilon_{BA}$ defined above, and is written, using the definition Eq. (\ref{firststar}) of the first star hBN basis $\vec g'_j$, as
\beqa \label{adhesion}
V_S(\vec r)&=&2\re\left[v_S\left(e^{i\vec g'_1\vec r}+e^{i\vec g'_2\vec r}+e^{i(-\vec g'_1-\vec g'_2)\vec r}\right)\right]+v_0\nonumber\\
&=&\sum^{\pm 3}_{j=\pm 1}v_je^{i\vec g'_j\vec r}+v_0
\eeqa
Vector $\vec r$ in Eq. (\ref{adhesion}) is the position, in the crystal plane, of the center of any given graphene unit cell, so that $\vec r=0$ corresponds to AA stacking, and $\vec r=\pm\vec r_0=
\pm(\vec a'_1+\vec a'_2)/3$ corresponds to AB/BA stacking. The complex numbers $v_j$ are defined as
\[
v_{j>0}=v_{j<0}^*=v_S=-\frac{\Delta\epsilon_{AB}+\Delta\epsilon_{BA}}{18}-i\frac{\Delta\epsilon_{AB}-\Delta\epsilon_{BA}}{6\sqrt{3}}
\]
so that the adhesion potential is a local extremum at these points, and their difference is indeed $\Delta\epsilon_{AB/BA}$. Note that the constant energy offset in Eq. (\ref{adhesion}),  $v_0=(\epsilon_{AB}+\epsilon_{BA}+\epsilon_{AA})/3$, is irrelevant for the purpose of computing the equilibrium deformations.

The total adhesion energy per graphene unit cell is the sum, over all $N$ graphene unit cells positions $\vec R_{\vec n}$ contained in a moir\'e supercell, of the adhesion potential
\[
U_S=\frac{1}{N}\sum_{\vec{n}}^N V_S(\vec R_{\vec n})
\]
The positions $\vec R_{\vec n}$ above are
\[
\vec R_{\vec n}=\vec r_{\vec n}+\vec u(\vec r_{\vec n})
\]
where
$
\vec r_{\vec n}=\vec n\bm a
$
are the unstrained unit cells positions, and $\vec n=(n_1,n_2)$ is a vector of integers.
Using the fact that $\bm g\bm a=2\pi$, we have $e^{i\vec g'_j\vec r_{\vec n}}=e^{i(\vec g'_j-\vec g_j)\vec r_{\vec n}}=e^{-i\vec G_j\vec r_{\vec n}}$, so that $U_S$ reads
\beqa
U_S&=&\frac{1}{\det \bm A}\int_{\bm A}d^2r \tilde V_S\left[\vec r,\vec u(\vec r)\right]\\
\tilde V_S\left[\vec r,\vec u(\vec r)\right]&=&\sum_{j=\pm 1}^{\pm 3}v_j e^{-i\vec G_j\vec r}e^{i\vec g'_j\vec u(\vec r)} \label{VStilde}
\eeqa
where we have transformed the sum into an integral over the moir\'e supercell, of area $\det \bm A$, since the form of the integrand $\tilde V_S\left[\vec r,\vec u(\vec r)\right]$ is now smooth on the atomic scale. This last step transforms our description into a continuum theory on the moir\'e supercell. Note however, that the large hBN momenta $\vec g'_j$ are retained, associated to the displacements $\vec u(\vec r)$.

To minimize the total energy analytically we need to assume that displacements $\vec u$ are small as compared to the hBN lattice constant. This is the linear distortion regime, and allows us to expand $\tilde V_S$ to first order in $\vec u(\vec r)$
\beq\label{uexp}
\tilde V_S\left[\vec r,\vec u(\vec r)\right]\approx \tilde V_S\left[\vec r,0\right]+\vec u(\vec r)\left.\partial_{\vec u}\tilde V_S\left[\vec r,\vec u(\vec r)\right]\right|_{\vec u=0}
\eeq
Using, once again, a harmonic decomposition for $\vec u(\vec r)$, Eq. (\ref{uharmonics}), we arrive at an adhesion energy that depends only on the harmonics $\vec u_{\vec q}$ for momenta $\vec q=\vec G_j$ in the first star of the moir\'e superlattice,
\[
U_S=i\sum_{j=\pm 1}^{\pm 3} v_j \vec g'_j \vec u_{\vec G_j}
\]
This is a generic feature of the linear distortion theory: if the microscopic adhesion profile $V_S(\vec r)$ is composed of a set of harmonics with momentum $\vec q_i=\vec m_i\bm g'$ (integer $\vec m_i$), the linearised adhesion energy will depend only on distortion harmonics with momentum $\vec m_i\bm G$.

The equilibrium value of distortion harmonics $\vec u_{\vec q}$ are obtained by minimising $U=U_S+U_E$. Since $U_E$ is quadratic on $\vec u_{\vec q}$, all harmonics \emph{different} from the $\vec u_{\vec G_j}$ in the adhesion energy will be zero in equilibrium. For the remaining six harmonics, we obtain, by differentiating $U$,
\beq \label{usol}
\vec u_{\vec G_j}=iv_j^*\bm{W}_{\vec G_j}^{-1}\vec g'_j
\eeq
This is the main analytical result of this section. At $\theta=0$, the $\vec u_{\vec G_j}$ become  $\vec u_{\vec G_j}=i[(1+\delta) v^*_j/(2 \delta^2 \pi B a_0^2)] \vec g'_j/|\vec g'_j|$.  We have checked that quadratic terms in the displacements, which arise from expanding the adhesion potential to second order, do not modify significantly this estimate. Moreover, the quadratic expansion confirm that the displacements in Eq. (\ref{usol}) are, at least, a local minimum of the sum of elastic and adhesion energies.

From Eq. (\ref{usol}) we can compute analytical expressions for a number of observables. In particular, one may compute the strain tensor $\bm u (\vec r)=u_{ij}(\vec r)=\frac{1}{2}[\partial_i u_j(\vec r)+\partial_j u_i(\vec r)]$, and other important related observables, such as the relative expansion of the lattice at a given point $\frac{1}{2}\mathrm{Tr}\bm u$. Evaluating e.g. the relative lattice expansion at $\theta=0$, we find a simple expression for the difference between the relative expansion in the center of the AB region and in AA regions
\beq \label{Tru}
\frac{1}{2}\Sigma_{i=x,y} [u_{i,i}^{AB} - u_{i,i}^{AA}]=\frac{\epsilon_\mathrm{AA}-\epsilon_\mathrm{AB}}{\sqrt{3}\delta  B a_0^2}
\eeq
This quantity has been measured to be greater than $2\%$.\cite{Woods:14, Matt:PC}


\subsection{Pseudogauge fields}

\begin{figure}
   \centering
   \includegraphics[width=0.32\columnwidth]{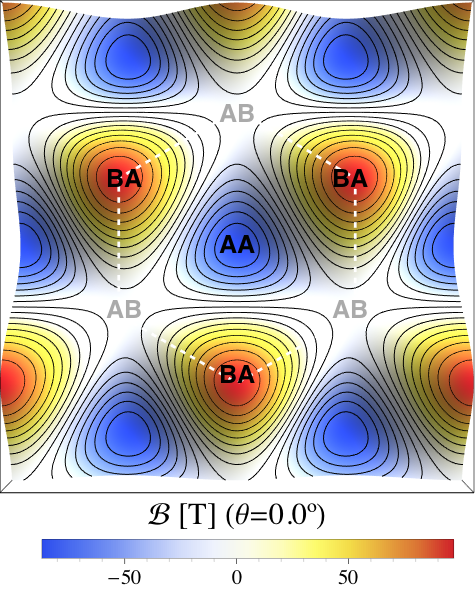}
   \includegraphics[width=0.32\columnwidth]{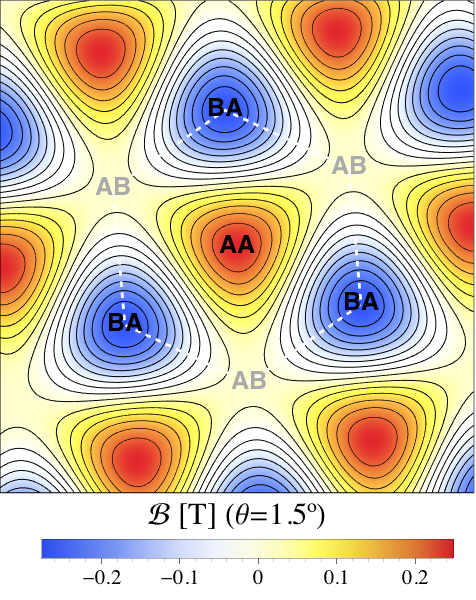}
   \includegraphics[width=0.32\columnwidth]{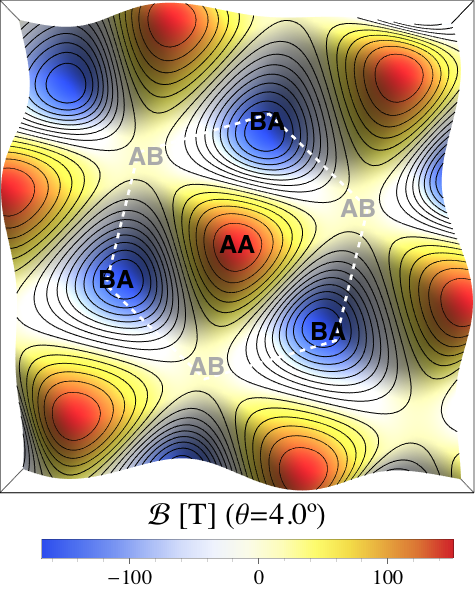}
   \caption{Pseudomangetic field $\mathcal{B}(\vec r)$ in real space for rotation angle $\theta=0$ (left), $\theta=1.5^\circ$ (center) and $\theta=4^\circ$ (right). Large fields above 200 T are produced by the strains.}
   \label{fig:B}
\end{figure}

\begin{figure}
   \centering
   \includegraphics[width=0.75\columnwidth]{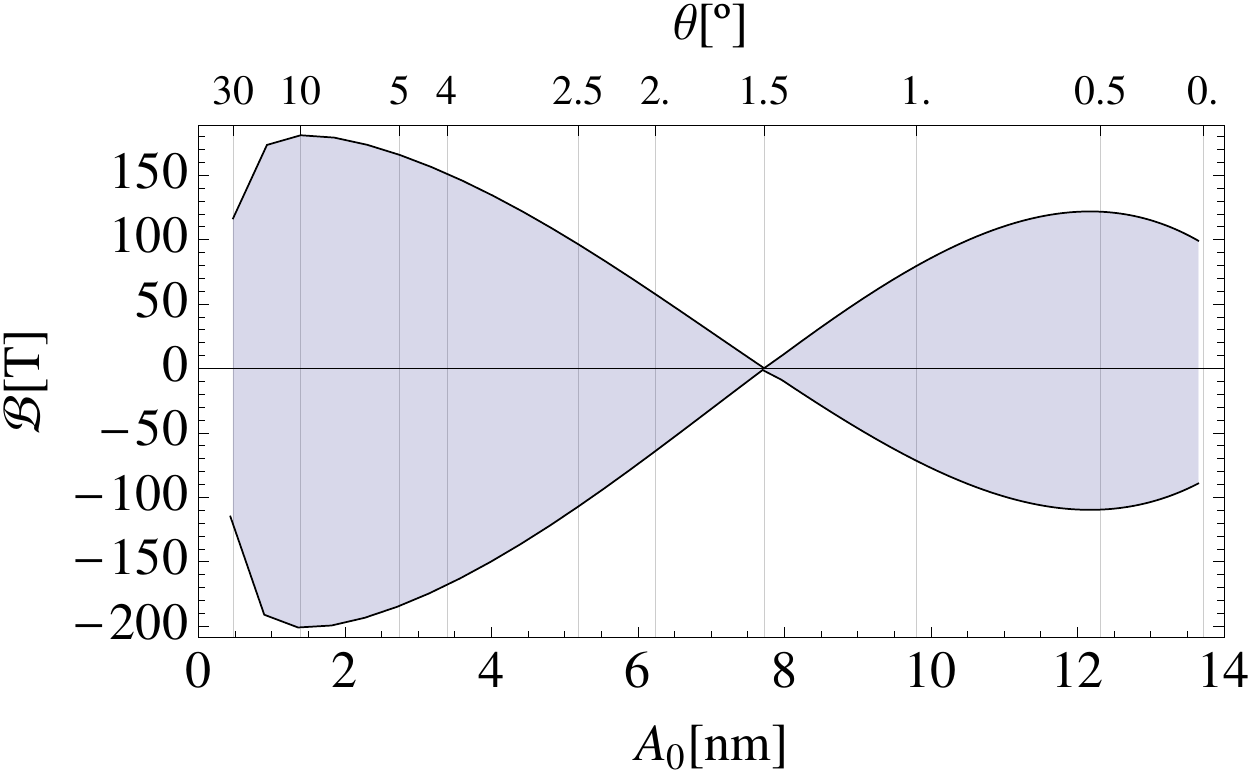}
   \caption{Range of variation of magnetic field $\mathcal{B}$ throughout the sample as a function of rotation angle $\theta$, or moir\'e period $A_0$. Note the large $\sim$ 200 T maximum fields, even for large angles, and the zero at $\theta_\mathcal{B}\approx 1.5^\circ$.}
   \label{fig:Brange}
\end{figure}

A strain field in graphene is known to produce an effective pseudogauge field, due to the modulation of nearest neighbor hopping amplitude $t\approx 2.78$ eV with the displacements \cite{Neto:RMP09}. In terms of the dimensionless parameter $\beta=d \log t/d \log a_0\approx 2$, the pseudogauge potential is given by
\beq
\label{A}
\vec{\mathcal{A}}(\vec r)=\pm\frac{\beta t}{e v_F}\left(\begin{array}{cc}
u_{xx}-u_{yy}\\-2u_{xy}
\end{array}\right),
\eeq
where the strain tensor $\bm u(\vec r)=u_{ij}(\vec r)$ is written in a coordinate frame with AB bond aligned along the $y$ direction, and the $\pm$ sign correspond to each of the two valleys (we will specialize on the $+$ sector in the follows, the opposite one trivially related by time reversal symmetry). In the next section we will analyse the effect of this field on the low energy electronic structure.

If we consider, conf. the solution Eq. \ref{usol}, that only the first star harmonics of $\vec u(\vec r)$ are non-zero, we obtain a pseudogauge potential that is likewise within the first star, $\vec{\mathcal{A}}(\vec r)=\sum_{j=\pm1}^{\pm 3} \vec{\mathcal{A}}_j e^{i\vec G_j\vec r}$, where the $\vec{\mathcal{A}}_j=\vec{\mathcal{A}}_{-j}^*$ harmonics can be written, conf. Eq. (\ref{A}), as
\beq
\vec{\mathcal{A}}_j=\frac{\beta t}{e v_F}
\left(\begin{array}{rr}
i\vec G_j\bm \sigma_z\vec u_j\\-i\vec G_j\bm \sigma_x\vec u_j
\end{array}\right),
\eeq
where  $\bm \sigma_i$ are Pauli matrices. The associated pseudomagnetic field $\mathcal{B}(\vec r)=\partial_x\mathcal{A}_y-\partial_y\mathcal{A}_x=\sum_{j=\pm1}^{\pm 3}\mathcal{B}_j
 e^{i\vec G_j\vec r}$ has harmonics $\mathcal{B}_j=\mathcal{B}_{-j}^*=-\vec G_j\bm \sigma_y\vec{\mathcal{A}}_j$.

It is interesting to note that while the typical equilibrium distortions of Eq. (\ref{usol}) scale as $A_0^2$ (since $\bm W_{\vec G_j}\sim A_0^{-2}$), the pseudomagnetic field $\mathcal{B}$ contains two spatial derivatives that cancel this scaling, so, unlike $\vec u$, it is not expected to vanish as the angle $\theta$ increases. Its effect on the electronic structure, however, will be diminished, since the physically relevant ratio of magnetic length to moir\'e period will increase.

For the case when the hBN and graphene axes are aligned, we can use the estimate for the strain in Eq. (\ref{Tru}), and obtain a typical value for the effective magnetic length $\ell_{\mathcal{B}}=\sqrt{\hbar/|e\mathcal{B}|}$ in terms of the elastic properties of graphene and the adhesion to the substrate
\begin{align}
\ell_{\mathcal{B}} &=\frac{3}{4\sqrt{\pi}}  \sqrt{\frac{\hbar(1+\delta)B a_0^2}{\beta | \epsilon_{AA} - \epsilon_{AB} |}}a_0
\label{beff}
\end{align}

\section{Discussion}\label{sec:discussion}

\begin{figure}
   \centering
   \includegraphics[width=0.32\columnwidth]{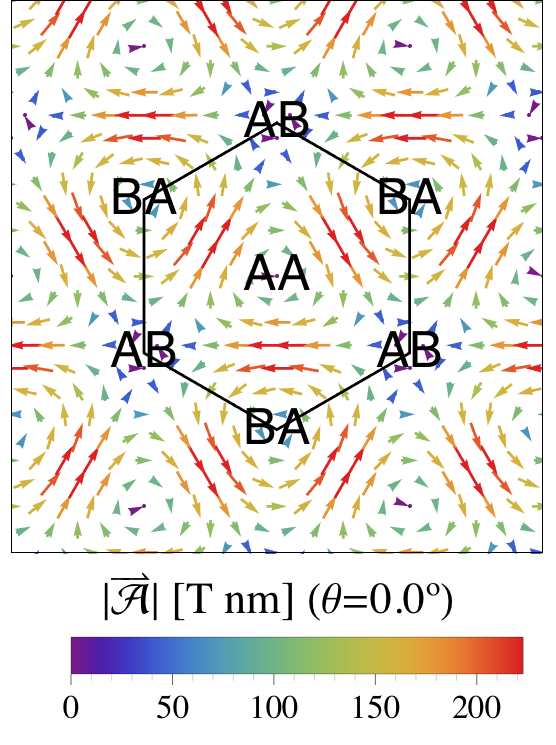}
   \includegraphics[width=0.32\columnwidth]{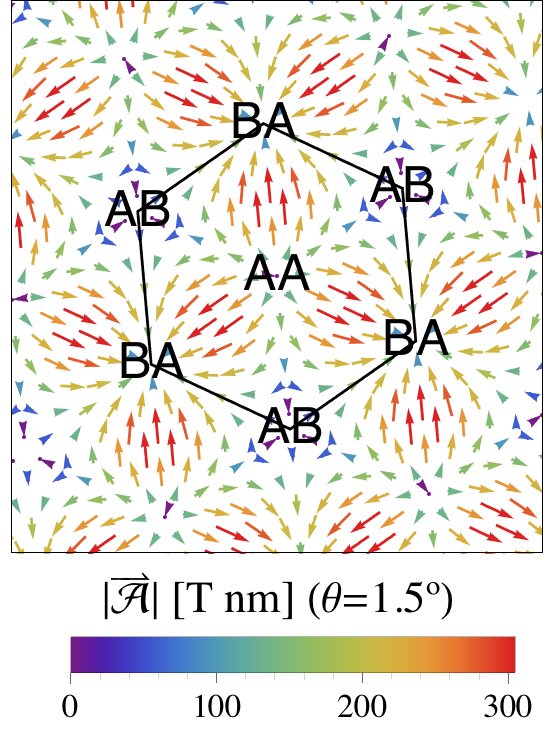}
   \includegraphics[width=0.32\columnwidth]{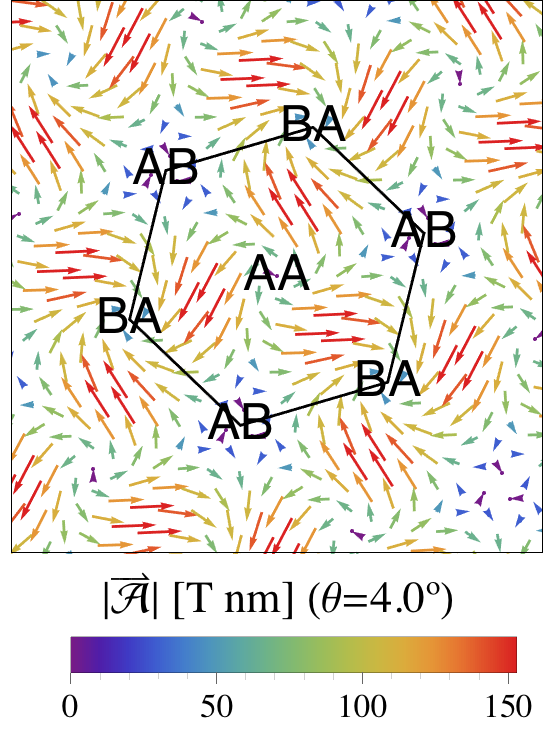}
   \caption{Pseudogauge potential $\vec{\mathcal{A}}(\vec r)$ in real space for rotation angle $\theta=0$ (left), $\theta=1.5^\circ$ (center) and $\theta=4^\circ$ (right). Note the vorticity inversion at $\theta=\theta_\mathcal{B}=1.5^\circ$.}
   \label{fig:A}
\end{figure}

The different quantities computed in the preceding section depend critically on the adhesion energy differences $\Delta\epsilon_{AB/BA}$, as compared to the typical elastic energy of graphene $\sim a_0^2 B\approx 97$ eV, multiplied by some power of $\delta\approx 1.8\%$ (recall that $\lambda\approx 3.5$ eV/\AA${}^2$ and $\mu\approx 7.8$ eV/\AA${}^2$). The adhesion energies have been computed using different ab-initio and semi-empirical approaches \cite{Sachs:PRB11,Caciuc:JPCM12,Berland:PRB13,Neek-Amal:14}. These calculations give values in the range of some tens of meV per unit cell for $\Delta\epsilon_{AB}$, and much lower for $\Delta\epsilon_{BA}$. On the other hand, the experiment of Ref. \onlinecite{Woods:14} has observed a difference of at least 2\% in the local lattice parameter between AB and AA regions. Using Eq. (\ref{Tru}), we see that, if the elastic moduli of graphene are not significantly modified by the presence of hBN, the adhesion energy differences should be at least -60 meV/unit cell to account for the observed deformation, with Refs. \onlinecite{Caciuc:JPCM12,Berland:PRB13} suggesting values even greater than -100 meV/unit cell when taking into account London dispersion forces. We use this latter value for $\Delta \epsilon_{AB}$, with $\Delta \epsilon_{BA}$ a tenth of that, which yields results in good agreement with the experiment.

The solution for the strain field of Eq. (\ref{usol}) is plotted in Fig. \ref{fig:u} for rotation angles $\theta=0^\circ$, $\theta=1.5^\circ$ and $\theta=4^\circ$. We see that the magnitude of the displacements is indeed much smaller than the lattice constant $a_0'=0.251$ nm, which justifies our linear expansion in $\vec g'_j\vec u(\vec r)$. We also see that at $\theta=0$, the solution approaches a pure longitudinal mode, that is even respect to the AB point, see Fig. \ref{fig:utypes}. This solution is thus dominated by local expansion.
As the angle is increased, we see how the solution crosses over to an even-transverse mode respect to the AB point, which is dominated by local twists and increased shear.
The local expansion $\frac{1}{2}\mathrm{Tr} \bm u$ associated to these distortions is shown in Fig. \ref{fig:Tru}. The equilibrium strain for the adhesion and elasticity parameters used reaches very large values for $\theta=0$. In the AB region the lattice expands by $da_0$, so that $a_0+da_0\approx a_0'$. The relative expansion $da_0/a_0=\frac{1}{2}\mathrm{Tr}\bm u$ reaches its maximum value $\delta=1.8\%$, as corresponds to adhesion dominating the total energy. In the other regions the lattice compresses by a comparable, though somewhat smaller amount, so that the difference surpasses 2\%, as found experimentally \cite{Woods:14}.


The adhesion energy $\tilde V_S[\vec r,\vec u(\vec r)]$ is shown in Fig. \ref{fig:Ad}. The flat blue regions around $\theta=0$ (left panel) correspond to AB regions in near-perfect registry, where the lattice locally expands by the effect of the adhesion. Surrounding these flat regions are sharp hexagonal boundaries, with (different) local maxima at the AA and BA points. It is clear that as the rotation angle $\theta$ increases and the moir\'e period decreases, the adhesion energy loses to the elastic energy, and the strain field is quickly suppressed. One way to quantify this effect is to analyse the $\theta$-dependence of the full width at half maximum (FWHM) of the adhesion potential as one moves between an AB region to the next. This is plotted in Fig. \ref{fig:FWHM}. A purely unstrained bilayer has a FWHM$=A_0/2$. We can see how this value decreases as $A_0$ is increased.

\begin{figure}
   \centering
   \includegraphics[width=0.9\columnwidth]{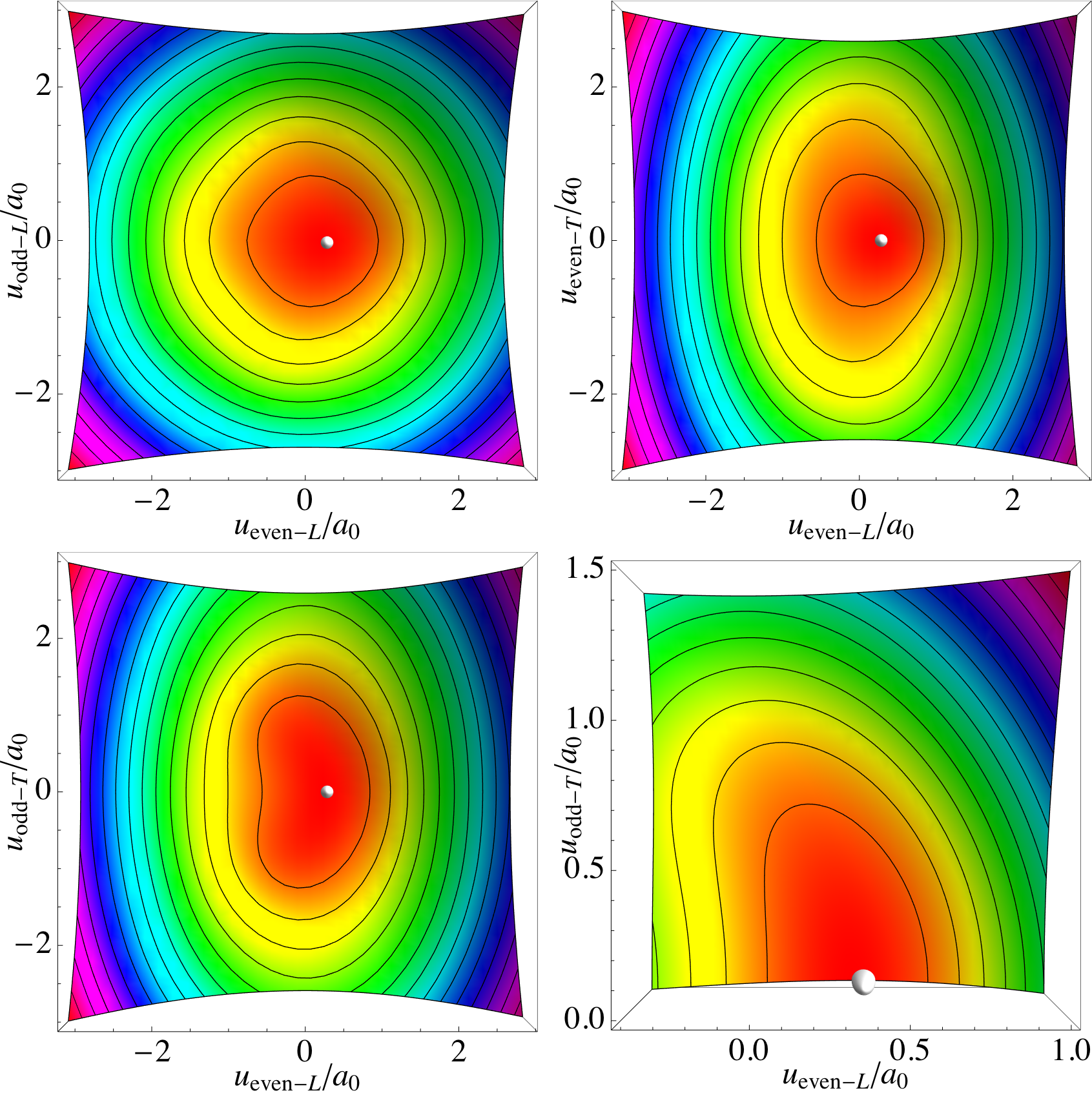}
   \caption{Total adhesion at $\theta=0$ beyond the linear approximation, as a function of pure distortion amplitudes $u_{\mathrm{even}/\mathrm{odd}-\mathrm{L}/\mathrm{T}}$, see Fig. \ref{fig:utypes}. Note that the minimum (white dot), given by Eq. (\ref{usol}), is absolute, and is not destabilised by non-linear corrections}
   \label{fig:potential}
\end{figure}

The spatial patterns of vertical Young modulus recently measured with atomic force microscopy (AFM) by Woods et al.  \cite{Woods:14} are strongly reminiscent of the adhesion potential profiles shown in Fig. \ref{fig:Ad}, including the small difference between AA and BA vertices along the hexagonal boundary (which are due to the finite $\Delta\epsilon_{BA}<0$). It can be argued that the measured elastic modulus should indeed reflect, to certain extent, the spatial modulation of the adhesion potential, since a stronger adhesion should correlate to a stiffer elastic modulus respect to vertical deformations. The FWHM of the experimental elastic modulus also shows a strong decrease as the angle approaches zero. However, the way this decrease happens is far more abrupt in the experiment than in our model, apparently dropping discontinuously at around $A_0\approx 10$ nm ($\theta=1^\circ$). This suggests effects beyond our present model, such as the possibility of additional contribution to the total energy, the formation of ripples whereby the interlayer distance acquires a spatial texture, or even a global commensurate-incommensurate transition, associated to a sudden jump in the area of the graphene sample as the angle is decreased.\cite{Chaikin:00} These considerations remain beyond the scope of this work, and require numerical computation of a rather different kind. We have evaluated within our analytical framework the effect of including additional harmonics to the adhesion potential in Eq. (\ref{adhesion}), as those described in Ref. \onlinecite{Neek-Amal:14}, but the results of Fig. \ref{fig:FWHM} do not change qualitatively. We have likewise excluded the possibility of a first- and second-order phase transitions as a result of non-linear terms in Eq. (\ref{uexp}). This is clear from the profile of the total energy $U$ around $\theta=0$, shown in Fig. \ref{fig:potential} as a function of longitudinal/transverse and even/odd distortion amplitudes, $u_{\mathrm{even}/\mathrm{odd}-\mathrm{L}/\mathrm{T}}$. Note that the potential minimum in Eq. (\ref{usol}) (white dot in the figure) remains stable and is the true absolute minimum of the potential. This remains valid even in the unrealistic extreme of vanishing shear modulus (not shown).

Finally, the pseudomagnetic field associated to the strain is shown in Fig. \ref{fig:B}. The large strains involved in the equilibrium configuration produce very large pseudomagnetic fields up to 200 T.  Surprisingly, however, the spatial pattern experiences an inversion at a finite but small angle $\theta_\mathcal{B}$, around which the pseudomagnetic field is suppressed and changes sign. The range of spatial variation of $\mathcal{B}$ as a function of $\theta$ is shown in Fig. \ref{fig:Brange}, which reveals the inversion at $\theta_\mathcal{B}\approx 1.5^\circ$. Analysing the vector potential $\vec{\mathcal{A}}(\vec r)$ at this particular rotation angle, we find a similar pattern as that in Fig. \ref{fig:u}(left): while $\vec{\mathcal{A}}(\vec r)$ is non-zero, it has a vanishing curl, so it is a pure gauge (purely longitudinal field, odd respect AB, see Fig. \ref{fig:utypes}). It's vorticity, in fact, becomes inverted at this $\theta_\mathcal{B}$. This is shown in Fig. \ref{fig:A}. Apart from the field inversion, the typical magnitude of the pseudomagnetic field is roughly in the 100-200 T throughout all angles, although its physical effects on the electronic structure should be stronger at small angles, where the magnetic length greatly exceeds the moir\'e period.

\section{Spectral gap}\label{sec:gap}

\begin{figure}
   \centering
   \includegraphics[width=0.8\columnwidth]{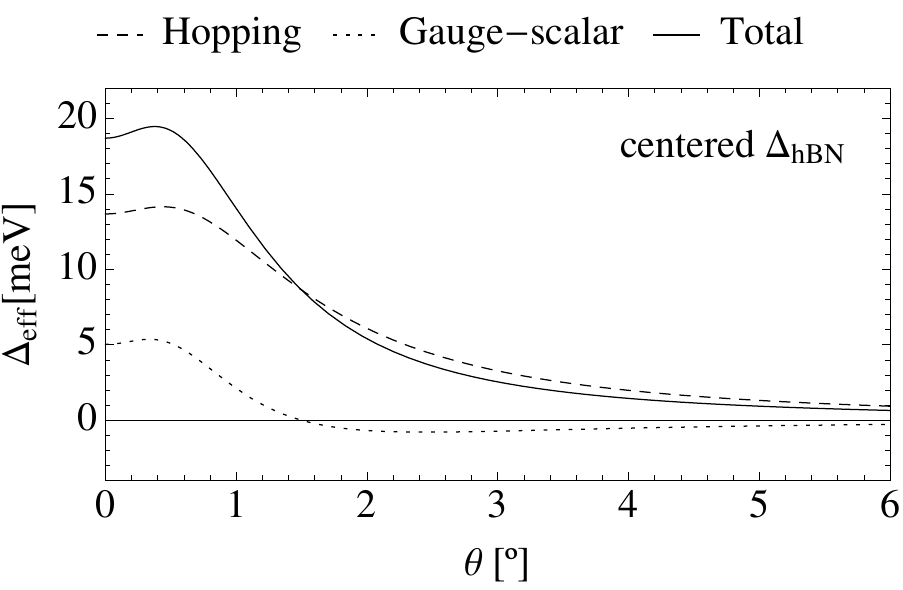}\\
      \includegraphics[width=0.8\columnwidth]{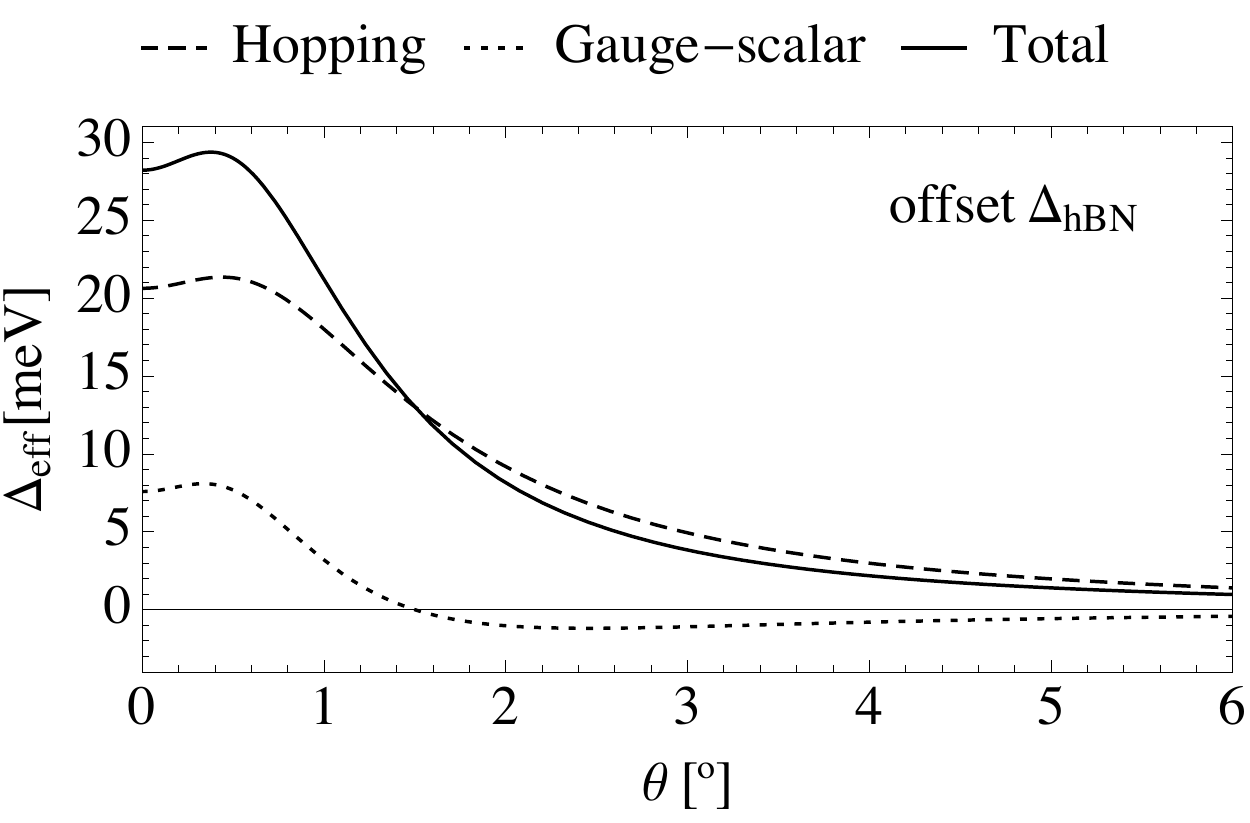}
   \caption{Effective spectral gap induced by hBN on graphene under equilibrium strains, as a function of relative angle $\theta$. We assume an hBN gap centered around graphene's neutrality point (top panel) and a 1.3 eV offset between the two (bottom panel), and decompose the gap into its two leading contributions (dashed and dotted line).}
   \label{fig:gap}
\end{figure}

The problem of assessing the spectral gap of graphene coupled to the gapped hBN crystal can be analysed assuming that the hBN gap ($\Delta_\mathrm{hBN}\approx 5.2$ eV) is much larger than the energy scales under consideration. In this limit, its effect on graphene's low energy effective Dirac Hamiltonian is the addition of a position dependent SU(2) self-energy $\bm \Sigma^{(0)}(\vec r)$. Its absolute magnitude is $m_0=t_\perp^2/(\Delta_\mathrm{hBN}/2)\approx 35$ meV, where $t_\perp\sim 0.3 eV$ is the graphene-hBN hopping amplitude. The gap at the Dirac point can be approximated to first order in $m_0$ as the spartial average average in a supercell $\bm A$ of area $\mathrm{det}\bm A$
\beq \label{Deltaeff}
\Delta_\mathrm{eff}=\frac{1}{\mathrm{det}\bm A}\int_{\bm A} d^2r \mathrm{Tr}\left[\bm\sigma_z\bm\Sigma^{(0)}(\vec r)\right]+\mathcal{O}(m_0^2)
\eeq

In the absence of strains, it can be shown that the local gap $\Delta(\vec r)=\mathrm{Tr}\left[\bm\sigma_z\bm\Sigma^{(0)}(\vec r)\right]$ has a zero average, so that the average gap is zero. The AB and BA regions will have a positive local gap $\Delta(\vec r_\mathrm{AB})=\Delta(\vec r_\mathrm{BA})=m_0$, while the AA region will have a negative local gap that exactly cancels the former, $\Delta(\vec r_\mathrm{AA})=-2m_0$.

The effect of spontaneous strains, as we saw, is to expand the AB regions at the expense of BA and AA. This breaks the cancelation of the average $\Delta (\vec r)$, and hence, strains will create a gap $\Delta_\mathrm{eff}\neq 0$ at the Dirac point. In the extreme case that the effective AB-stacked area grows from $A/3$ to cover most of the supercell area $A$, the average gap will become $m_0\approx 35$ meV. In the more realistic case described here, the linear size of the AB region at $\theta=0$ is around $70\%-75\%$ of the supercell diameter, which yields an estimate for the gap around $15$ meV. A figure closer to the maximum $m_0$ would be obtained for stronger adhesion parameters, which would result in a larger AB region (smaller FWHM in Fig. \ref{fig:FWHM}).

Interestingly, it has been noted \cite{Low:PRB11} that in the presence of strains, there is another contribution to Eq. (\ref{Deltaeff}) that further increases the effective gap by around 40\% [actually $\beta/(\sqrt{3}\pi)$, to be precise]. This comes about in second order of perturbation theory in the pseudogauge field $\vec{\mathcal{A}}(\vec r)$ and the scalar potential $\mathrm{Tr}\left[\bm\sigma_0\bm\Sigma^{(0)}(\vec r)\right]$. Crucially, both the $\sim m_0$ term in the preceding paragraph and this second order contribution are parametrically equal in a systematic expansion in the deformations $\vec u$ and the inverse hBN gap $\Delta_\mathrm{hBN}^{-1}$. A careful evaluation of the two contribution yields at $\theta=0$,
\beqa
\label{Delta}
\Delta_\mathrm{eff}&=&\frac{2}{\sqrt{3}}
\frac{1+\delta}{\delta^2}\left(1+\frac{\beta}{\sqrt{3}\pi}\right)\\
&&\times\frac{|m_-(\Delta\epsilon_{BA}-\Delta\epsilon_{AB})+m_+(\Delta\epsilon_{BA}+\Delta\epsilon_{AB})|}{9 a_0^2 B}\nonumber
\eeqa
Here, $m_\pm=\frac{t_\perp^2}{2}(\epsilon_c^{-1}\pm\epsilon_v^{-1})$ is given in terms of the conduction and valence band edges $\epsilon_{c,v}$ in hBN respect to graphene's neutrality point (if the gap is centered, $m_-=m_0$, and $m_+=0$).   Recall also that $\delta\approx 1.8\%$, $\beta\approx 2$, and $B\approx 19.1$ meV/\AA${}^2$ is the bulk modulus of graphene.  With our assumption for the adhesion energies, the gap at $\theta=0$ is approximately $\Delta_\mathrm{eff}\approx20$ meV.
This value for $\Delta_\mathrm{eff}$ is in qualitative agreement with experimental observations \cite{Hunt:S13}.
The effective gap $\Delta_\mathrm{eff}$ as a function of $\theta$ is shown in Fig. \ref{fig:gap}. Note that the gauge-scalar contribution vanishes, as expected, at the special $\theta_\mathcal{B}\sim 1.5^\circ$ angle, for which the pseudomagnetic field vanishes.
Note that a finite energy offset between the Dirac point and the gap center of hBN ($m_+\neq 0$) can result in a further increase of the induced gap. As an example, a shift of $\sim 1.3$ eV between the two yields a $\theta=0$ value $\Delta_\mathrm{eff}\approx 30$ meV (see Fig.\ref{fig:gap}, bottom panel), in quantitative agreement with experiment.

\section{Conclusions}\label{sec:conclusion}

We have presented a model for the in plane deformations of a graphene layer on a hBN substrate.
The deformations, effective magnetic field, and average gap can be expressed in terms of the elastic properties of graphene, the lattice mismatch, and the adhesion energy between graphene and hBN, see Eqs. (\ref{Tru}), (\ref{beff}), and (\ref{Delta}). The estimates presented here give an electronic gap of a few tens of meV, in line with experiments.
The average strains near perfect alignment are a few percent $\sim \delta$, and give rise to effective pseudomagnetic fields of order 50-100T.  The pseudomagnetic length is a few nanometers, about one order of magnitude smaller than the dimensions of the superlattice unit cell, which should therefore lead to strong effects in the electronic structure.

The different components of the potential induced by the moir\'e superlattice include even and odd terms under spatial inversion of similar magnitude, as expected from an hBN substrate. The combination of a modulated scalar and gauge potential gives a contribution to the average gap which has the same parametric dependence and order of magnitude as the gap arising from the enlargement of the energetically favorable $AB$ regions.

The main results arise from a competition between the rigidity of the graphene layer and the adhesion potential provided by the substrate. For realistic parameters, the graphene deformations are small, and pinning and commensuration effects are not important. In terms of an effective Frenkel-Kontorova model, the results presented here are consistent with a floating phase, with gapless acoustic modes.


\section{Acknowledgements}
We thank K. S. Novoselov, A. K. Geim, A. Cortijo and F. Barbero for useful discussions. We acknowledge support from the Spanish Ministry of Economy (MINECO)
through Grant Nos. FIS2011-23713 and PIB2010BZ-00512, the European Research Council Advanced Grant (contract 290846), and
the European Commission under the Graphene Flagship,
contract CNECT-ICT-604391.
\bibliography{biblio}
\end{document}